\documentclass[a4paper,10pt, twoside]{cpc-hepnp}
\usepackage{multicol}
\usepackage{graphicx}
\usepackage{booktabs}
\usepackage{amssymb,bm,mathrsfs,bbm,amscd}
\usepackage[tbtags]{amsmath}
\usepackage{lastpage}

\begin{document}
\fancyhead[c]{\small Chinese Physics C~~~Vol. XX, No. X (201X)
    XXXXXX} \fancyfoot[C]{\small 010201-\thepage}

\title{On the knee of Galactic cosmic rays in light of sub-TeV spectral hardenings}
\author{Yi-Qing Guo$^{1,2}$ and Qiang Yuan$^{1,3}$}
\maketitle
\address{
$^1$Key Laboratory of Dark Matter and Space Astronomy, Purple Mountain
Observatory, Chinese Academy of Sciences, Nanjing 210008, Jiangsu, China;
yuanq@pmo.ac.cn\\
$^2$Key Laboratory of Particle Astrophysics, Institute of High Energy
Physics, Chinese Academy of Sciences, Beijing 100049, China;\\
$^3$School of Astronomy and Space Science, University of Science and 
Technology of China, Hefei 230026, Anhui, China}

\begin{abstract}
More than fifty years after the discovery of the knee in the cosmic ray 
(CR) spectra, its physical origin remains a mystery. This is partly due 
to the ambiguity of the energy spectrum of individual composition. 
Recently, direct measurements from several space experiments found 
significant spectral hardenings of CR nuclei at $\sim 200$ GV. 
A joint modeling of the direct and indirect measurements may help to 
understand the experimental systematics and probably the physics of the 
knee. In this work, we update the phenomenological ``poly-gonato'' model 
to include the spectral hardenings, with a changing spectral index of 
$\gamma + \beta \cdot \log E$. This modification gives a reasonable 
description to the CR spectra in a wide energy range. However,
the fits to different data sets result in somehow different results.
We find that the fit to the AMS-02 and CREAM data slightly favors a 
relatively low energy knee of the light components. In such a case, 
the expected all-particle spectra under-shoot the data, which may require 
an extra component of CRs. The fits to AMS-02 data and the light component 
(H+He) data from Tibet AS$\gamma$/ARGO-YBJ/WFCTA or KASCADE experiments 
give consistent results with the all-particle spectra. 
 We further propose a possible physical realization of such a ``modified 
poly-gonato'' model of spectral hardenings by means of spatially-dependent 
diffusion of CRs. We find reasonably good agreement between the model
predictions and the data about CR spectra, the secondary-to-primary 
ratios, and the amplitude of anisotropies.

\end{abstract}

\begin{multicols}{2}
\section{Introduction}

Nearly sixty years after the discovery of the knee in the CR 
spectra \citep{KK1958}, its underlying physical mechanism is still under
debate \citep{2004APh....21..241H}. It is generally accepted that each
composition has its own knee and the superposition of all compositions
gives the observed break of the all-particle spectra at $\sim4$ PeV.
This is the so-called ``poly-gonato'' model \citep{2003APh....19..193H}.
The energy of the knee of each composition may be proportional to charge
($Z$-dependent) or atomic number ($A$-dependent), which can be used to
probe the physical mechanism of the knee \citep{2004APh....21..241H}.
For example, the acceleration limit or propagation leakage may predict a
$Z$-dependence of the knee of each composition \citep{1983A&A...125..249L,
1988ApJ...333L..65V,1993A&A...268..726P,1996APh.....5..367B,
1998A&A...330..389W}. On the other hand, an $A$-dependence may imply 
an interaction origin of the knee, such as the photo-disintergration 
\citep{1993APh.....1..229K,2002APh....17...23C}, and inelastic collisions 
between CRs and background photons or neutrinos \citep{2001ICRC....5.1760K,
2009ApJ...700L.170H,2010SCPMA..53..842W,2013NJPh...15a3053G}.

The energy spectrum of individual nuclei composition is crucial to
understand the knee puzzle. Many efforts have been paid to measure the
individual spectrum with air shower experiments, however, no consensus
has been achieved yet, primarily due to the systematical uncertainties
of the absolute energy calibration. Some progresses in the spectral
measurements of individual composition in PeV energies have been made
in recent years with ground-based experiments. Although these measurements
themselves are not completely consistent with each other, they may
commonly suggest a knee below PeV for the light components
\citep{KneeHuangJing2013,2016PhRvL.117o1103A, 2013APh....47...54A,2015PhRvD..91k2017B,
2015PhRvD..92i2005B,2015arXiv150204840D,2016arXiv160801389M}. 
Compared with the $\sim4$ PeV knee of the all-particle spectra, such a 
result indicates that the knee is dominated by nuclei heavier than Helium
\citep{2010ApJ...716.1076S,2015ChPhC..39l5001Z,2016A&A...595A..33T}.

The direct measurements of lower energy CRs by balloon-borne or space
detectors can determine the individual spectrum much better, which were then
extrapolated to high energies to bridge the direct and air shower experiments
\citep{2003APh....19..193H,2010ApJ...716.1076S,2015ChPhC..39l5001Z}.
The extrapolation was usually based on power-law fits to the low energy
data. However, remarkable spectral hardenings at an rigidity of a few
hundred GV on the spectra of all major nuclei components were reported
by the balloon-borne experiments ATIC-2 \citep{2007BRASP..71..494P} and
CREAM \citep{2010ApJ...714L..89A}, and were confirmed with higher precision
by the space detectors PAMELA \citep{2011Sci...332...69A} and AMS-02
\citep{2015PhRvL.114q1103A,2015PhRvL.115u1101A}. Many kinds of models
have been proposed to understand the origin of the spectral hardenings,
including the super-position of different sources
\citep{2006A&A...458....1Z,2011PhRvD..84d3002Y,2012MNRAS.421.1209T},
the non-linear acceleration of the supernova remnant shocks
\citep{2010ApJ...725..184B,2013ApJ...763...47P}, the re-acceleration
mechanism when particles propagating in the Galaxy \citep{2014A&A...567A..33T},
and the spatially-dependent diffusion of CRs \citep{2012ApJ...752L..13T,
2015ApJ...815L..25G,2015PhRvD..92h1301T,2016ChPhC..40a5101J,
2014arXiv1412.8590G,2016ApJ...819...54G,2016PhRvD..94l3007F}.

Given these new measurements of both the direct and indirect experiments,
we re-visit the ``poly-gonato'' model of CRs in this work. We tend to build
an updated phenomenological model of the energy spectrum of each
composition which matches these newest data. We adopt a log-parabolic
spectrum with an asymptotically hardening spectral index of $\gamma+\beta
\cdot\log E$ to describe the spectral hardenings. An exponential cutoff
is employed to describe the knee of CRs. Through fitting to different
data sets with the two key parameters, $\beta$ and the cutoff energy
$E_c$, we further test the consistency among different measurements.
 One possible physical scenario for such a phenomenological 
``modified poly-gonato'' model is the spatially-dependent diffusion of 
CRs \citep{2012ApJ...752L..13T}. In such a model the propagation volume 
was divided into two regions, the inner halo and the outer halo. 
The key point is that the diffusion is slower and has a shallower 
rigidity-dependence in the inner halo than in the outer one, which 
can result in a break of the spectrum. As an illustration of a physical 
implementation of this ``modified poly-gonato'' model, we will also 
discuss this spatially-dependent diffusion scenario and compare its
predictions with the observational data.

\section{``Modified poly-gonato'' model}

\subsection{Model description}

The ``poly-gonato'' model to describe the knee is basically based on the
extrapolation of low-energy measurements. Up to the knee energies, typically
three types of models were employed to fit the all particle spectrum.
The first type is motivated by the diffusive shock acceleration or
propagation process. In those models, the cutoff energies of CRs are
expected to be proportional to the particle charge $Z$
\citep{1983A&A...125..249L,1993A&A...268..726P}. The second type is
motivated by interaction processes, in which the cutoff or break energies
are proportional to the atomic number $A$ \citep{1993APh.....1..229K,
2001ICRC....5.1760K,2002APh....17...23C,2009ApJ...700L.170H}.
The third type of the break is constant for all species. It is not well
physically motivated, but might be a simple assumption
\citep{2003APh....19..193H}. Recent results show that the break energy
of light components is lower than that of the all-particle knee, which
disfavors this constant break energy scenario \citep{KneeHuangJing2013,
2015PhRvD..91k2017B,2015PhRvD..92i2005B,2015arXiv150204840D}. Therefore,
only the $Z$-dependent and $A$-dependent cases are considered in the
following discussion.

To include the spectral hardenings at $\sim$200 GV, we parameterize the
spectrum of each composition as
\begin{eqnarray}
\label{log-parabola}
\frac {d\Phi^i}{dE}(E) & = &\Phi^i_0\times\left(\frac{E}{E_{\rm br}}\right)
^{-\gamma_1^i} \nonumber \\
&\times & \left (\frac{1+E/E_{\rm br}}{2}\right)^{\left [\gamma_1^i-
\gamma_2^i+\beta \cdot \log(E/E_{\rm had})\right]}\\
        & \times&  e^{-E/E^i_c}, \nonumber
\end{eqnarray}
where $E_{\rm br}$ is the break energy which describes the low energy (with
a rigidity of a few GV) behavior of the spectrum, $\Phi^i_0$ is the absolute
flux of the $i$th element at $E_{\rm br}$, $\gamma_1^i\,(\gamma_2^i)$
is the spectral index below (above) $E_{\rm br}$, $E_{\rm had}$ is the
energy characterizing the spectral hardening which is fixed to be 
$Z\cdot200$ GeV, and $\beta\cdot\log(E/E_{\rm had})$ is an asymptotically 
hardening term used to describe the spectral hardening.

The proton and Helium spectra have been measured up to TeV by AMS-02 
with very high precision \citep{2015PhRvL.114q1103A,2015PhRvL.115u1101A}.
Their spectral parameters are fitted separately with Eq.~(1). For the other 
major compositions, such as C, O, Mg, Al, Si, and Fe, the HEAO-3 data 
\citep{1990A&A...233...96E} are used to determine their spectral parameters. 
For simplicity, their low energy spectral parameters $\gamma_1$ and 
$R_{\rm br}\equiv E_{\rm br}/Z$ are assumed to be the same. For convenience, 
the spectral parameters of all species are tabulated in the Appendix.

To account for the spectra around the knee, we assume a $Z$- or 
$A$-dependent cutoff of each species as
\begin{equation}
E_c^i = \left \{
     \begin{array}{ll}
     E_c^p\cdot Z, & {\rm charge~dependent}\\
     E_c^p\cdot A, & {\rm mass~dependent}\\
     \end{array}
     \right .
\end{equation}
where $E_c^p$ is the cutoff energy of protons. Parameters $E_c^p$ correlates
with $\beta$. They will be determined through fitting to the data.

\subsection{Fitting results}

The direct measurements of proton and Helium fluxes by AMS-02
\citep{2015PhRvL.114q1103A,2015PhRvL.115u1101A} and CREAM I+III
\citep{2017ApJ...839....5Y}, as well as the air shower array 
measurements of the light components (H+He) at high energies 
\citep{KneeHuangJing2013,2015PhRvD..91k2017B,2015PhRvD..92i2005B,
2016arXiv160801389M} are used in the fits. The all-particle
spectra are not included in the fits. We require that the
calculated all-particle spectra are lower than the $2\sigma$ 
upper bounds of the observations. Due to the uncertainties of the 
absolute energy calibration and the hadronic interaction models 
of the ground based CR experiments, the observed break energies 
of the knee of light components differ from each other. 

\begin{figure*}[!htb]
\centering
\includegraphics[width=0.45\textwidth]{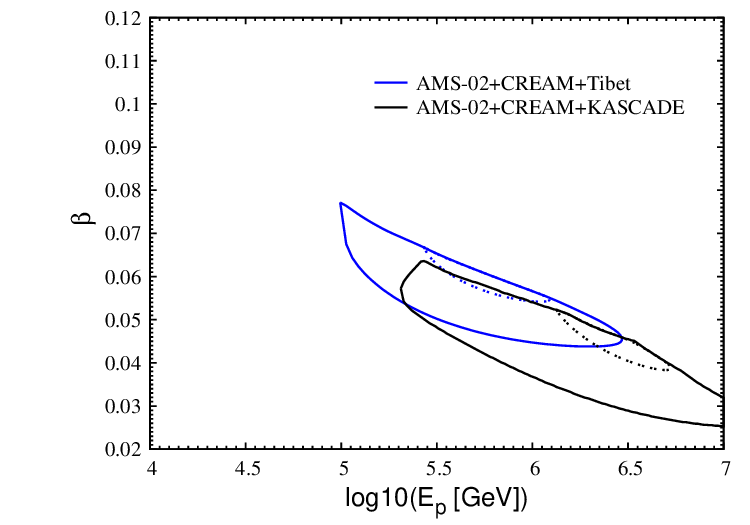}
\includegraphics[width=0.45\textwidth]{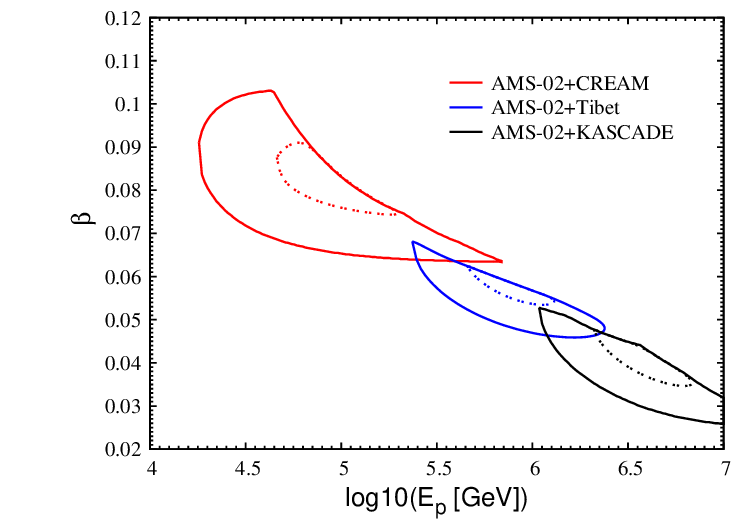}
\includegraphics[width=0.45\textwidth]{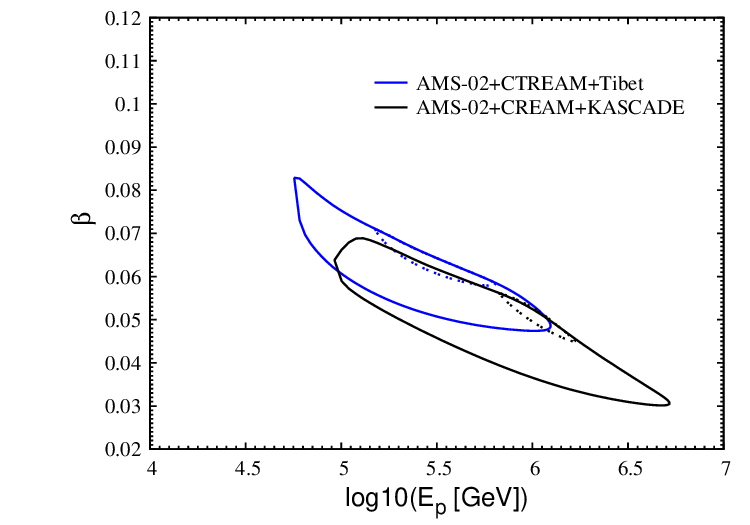}
\includegraphics[width=0.45\textwidth]{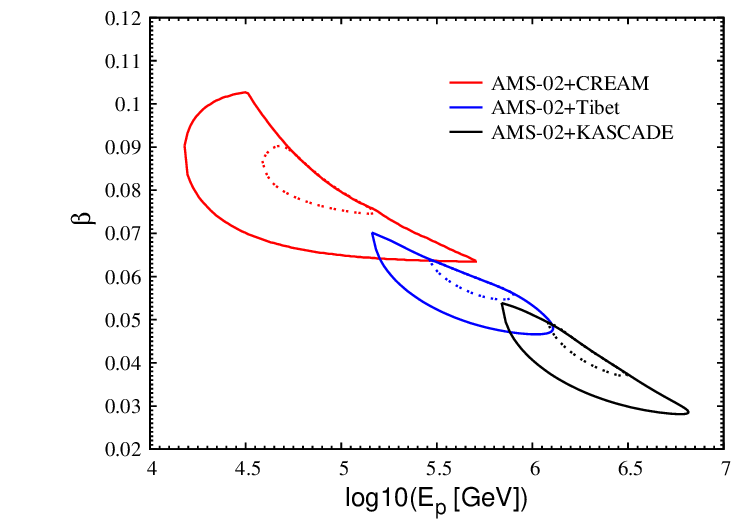}
\caption{The 68\% (inner dashed) and 95\% (outer solid) confidence 
regions of parameters $\beta$ and $E_c^p$. The left panels are for the
combined fits to AMS-02+CREAM+Tibet and AMS-02+CREAM+KASCADE, and the
right panels are for the fits to AMS-02+CREAM, AMS-02+Tibet, and 
AMS-02+KASCADE data, respectively. The upper panels are for Z-dependent 
and the lower panels are A-dependent cases.}
\label{fig:contourZ}
\end{figure*}
\small
\begin{table*}[!htb]
\begin{center}

\caption{Low energy spectral parameters of protons and Helium nuclei of 
the ``modified poly-gonato'' model}
\begin{tabular}{ccccccc}\hline \hline
\scriptsize 
Parameters                              & \tiny Species & \tiny AMS-02+CREAM+Tibet & \tiny AMS-02+CREAM+KASCADE & \tiny AMS-02+CREAM & \tiny AMS-02+Tibet & \tiny AMS-02+KASCADE \\ \hline
\normalsize
$\Phi_0$                                & p  & 337.6 & 383.0 & 290.6 & 351.3 & 371.4 \\ 
(m$^{-2}$s$^{-1}$sr$^{-1}$GeV$^{-1}$)   & He & 9.5   & 10.9  & 8.0   & 9.0   & 10.7  \\\hline
$R_{\rm br}$                            & p  & 2.15  & 1.94  & 2.41  & 2.07  & 1.99  \\
  (GV)                                  & He & 3.36  & 2.94  & 4.11  & 3.82  & 3.41  \\\hline
  $\gamma _1$                           & p  &$-0.63$&$-0.75$&$-0.49$&$-0.69$&$-0.70$\\
                                        & He & 0.07  & 0.00  & 0.15  & 0.12  & 0.02  \\\hline
  $\gamma _2$                           & p  & 2.93  & 2.93  & 2.96  & 2.93  & 2.92  \\
                                        & He & 2.84  & 2.83  & 2.86  & 2.84  & 2.82  \\
\hline \hline
\end{tabular}
\label{paramLow}
\end{center}
\end{table*}
\normalsize

We first classify the ground-based experiments into two groups, the 
Tibet group (including AS$\gamma$, ARGO-YBJ, and WFCTA) and KASCADE.
Together with the AMS-02 and CREAM data, we fit to each group of data
using both the $Z$- and $A$-dependent parameterizations of the knee.
The best-fit parameters are given in Tables \ref{paramLow} and
\ref{paramMPG}. The results show that the fit to AMS-02+CREAM+KASCADE
data gives a relatively large $\chi^2$ value, while the fit to
AMS-02+CREAM+Tibet data is acceptable. This is due to that the
KASCADE data favors a relatively high energy of the knee, which
requires $\beta$ to be relatively small and the spectral hardening 
effect is not enough to match the CREAM data (see Fig.~\ref{fig:contourZ}
for the contours of parameters $\log(E_p)$ and $\beta$\footnote{ An
anti-correlation between $\log(E_p)$ and $\beta$ is shown, which is
basically due to a mathematical constraint. For a larger $\beta$,
the spectra of individual compositions can not extend to very high 
energies without exceeding the all-particle spectra, and hence
$\log(E_p)$ is required to be smaller.}).
To compare with the standard poly-gonato model, we perform the 
fittings with $\beta=0$. The minimum $\chi^2$ values of these fittings 
are given in Table \ref{paramMPG}. It is obvious that these fittings 
are much worse than the ``modified poly-gonato'' model. 

\begin{figure*}[!htb]
\centering
\includegraphics[width=1\textwidth]{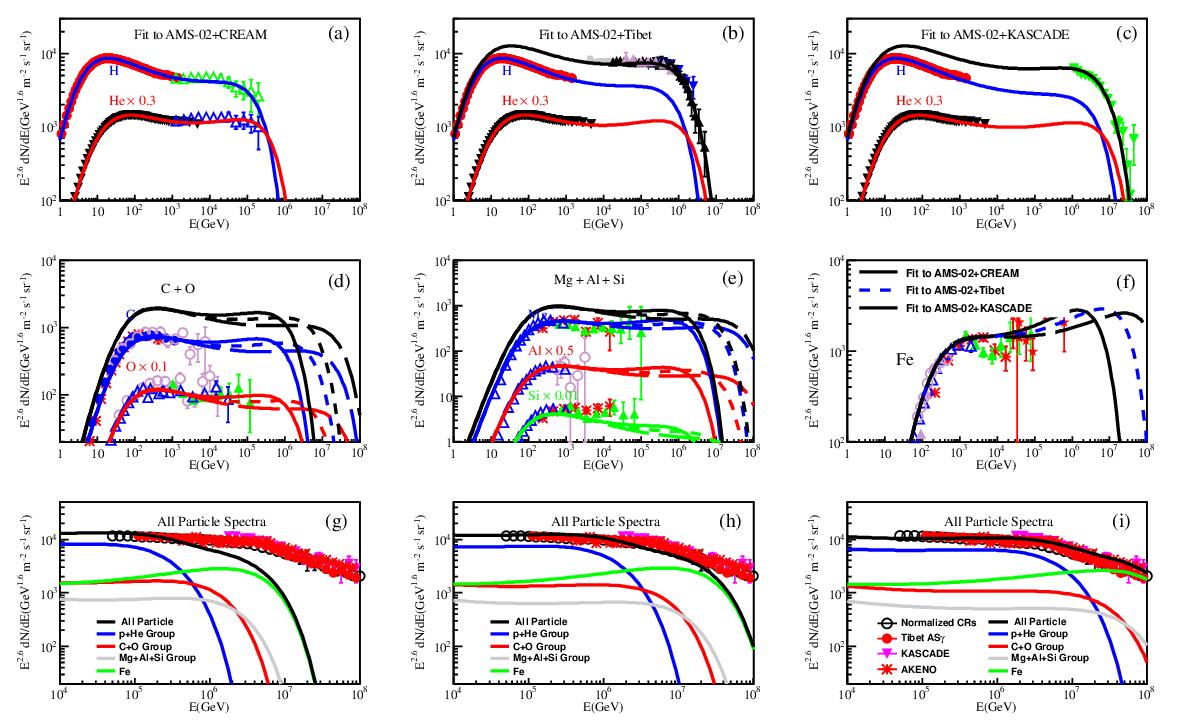}
\caption{The comparison between the best-fit results and the experimental
data, for the $Z$-dependent case. The proton data are from:
AMS-02 \citep{2015PhRvL.114q1103A}, CREAM \citep{2017ApJ...839....5Y},
ATIC-2 \citep{2007BRASP..71..494P}; the Helium data are from: AMS-02
\citep{2015PhRvL.115u1101A}, CREAM \citep{2017ApJ...839....5Y};
ATIC-2 \citep{2007BRASP..71..494P}; the Carbon, Oxygen, Magnesium, Aluminium,
Silicon, and Iron data are from: HEAO-3 \citep{1990A&A...233...96E}, TRACER \citep{2008ApJ...678..262A}, 
ATIC-2 \citep{2009BRASP..73..564P} and CREAM-II \citep{2009ApJ...707..593A};
the proton + Helium data are from: KASCADE \citep{2013APh....47...54A},
Tibet-AS$\gamma$ \citep{KneeHuangJing2013}, WFCTA \citep{2015PhRvD..92i2005B},
ARGO-YBJ \citep{2015PhRvD..91k2017B,2015arXiv150204840D}; the all-particle
data are from: Tibet-AS$\gamma$ \citep{2008ApJ...678.1165A}, KASCADE
\citep{2013APh....47...54A}, Akeno \citep{1984JPhG...10.1295N}, and
the normalized average one \citep{2003APh....19..193H}.}
\label{fig:logEZ}
\end{figure*}

Motivated by the facts that the CREAM data might reveal hint of
spectral softening above $\sim20$ TeV \citep{2017ApJ...839....5Y} and
the combined fit of AMS-02+CREAM+KASCADE does not give a good enough 
fit, we separate the CREAM data from the ground-based measurements
and re-do the fits with AMS-02+CREAM, AMS-02+Tibet, and AMS-02+KASCADE
data, respectively. The favored confidence regions of parameters 
$\log(E_p)$ and $\beta$ are shown in the right panel of 
Fig.~\ref{fig:contourZ}. It is shown that the AMS-02+CREAM fit
tends to favor a relatively low break energy of $E_p$ compared with
the other two fits. The parameter regions of AMS-02+Tibet and 
AMS-02+CREAM (AMS-02+KASCADE) overlap with each other at the 95\% 
confidence level. However, the results of AMS-02+CREAM and AMS-02+KASCADE
do not overlap.
\small
\begin{table*}[!htb]
\begin{center}
\caption{High energy spectral parameters of the ``modified poly-gonato'' model}
\begin{tabular}{ccccccc}\hline \hline
\tiny Parameters & \tiny Mode & \tiny AMS-02+CREAM+Tibet & \tiny AMS-02+CREAM+KASCADE & \tiny AMS-02+CREAM & \tiny AMS-02+Tibet & \tiny AMS-02+KASCADE \\
   \hline
   $\beta$ & $Z$     & 0.063 & 0.045 & 0.083 & 0.062 & 0.041 \\
           & $A$     & 0.068 & 0.051 & 0.083 & 0.065 & 0.048 \\\hline
   $E_c^p$ & $Z$     & $5.7\times10^5$ & $2.9\times10^6$ & $1.4\times10^5$ & $7.4\times10^5$ & $3.7\times10^6$ \\
   (GeV)   & $A$     & $2.9\times10^5$ & $1.3\times10^6$ & $1.3\times10^5$ & $4.1\times10^5$ & $1.7\times10^6$ \\\hline
   $\chi ^2$/dof  & $Z$  & 128.5/195 & 225.0/171 & 68.7/154 & 63.4/171 & 76.4/147 \\
   $\chi ^2$/dof  & $A$  & 116.7/195 & 191.1/171 & 68.5/154 & 71.7/171 & 71.6/147 \\
\hline 
 $\beta$=0  &     &  &  &  &  &  \\ \hline
 $\chi ^2$/dof  & $Z$  & 951.2/196 & 1042.8/172 & 678.4/155 & 653.2/172 & 399.9/148 \\
 $\chi ^2$/dof  & $A$  & 949.4/196 & 1001.6/172 & 678.2/155 & 651.7/172 & 399.8/148 \\
\hline \hline
\end{tabular} 
\label{paramMPG}
\end{center}
\end{table*}
\normalsize
Fig. \ref{fig:logEZ} shows the best-fit results of fluxes of several major 
components in CRs for the $Z$-dependent scenario, compared with the data. 
In this figure, panels (a)-(c) are the spectra of protons, Helium, and 
H+He for the three groups of fits (AMS-02+CREAM, AMS-02+Tibet, and
AMS-02+KASCADE). Panel (d) is for C and O, panel (e) is for Mg, Al, and 
Si, panel (f) is for Fe, respectively. Panels (g)-(i) are the all-particle 
spectra of the three groups. 

We find that for the AMS-02+CREAM fit, the favored energy of the knee 
of light components is relatively low, which under-shoots the all-particle 
spectra. However, given the large uncertainties and limited coverage of 
the energy range of the CREAM data, the constraint on the cutoff energy 
is very loose (see Fig.~\ref{fig:contourZ}). The Tibet experimental data 
gives a median knee energy of light components and is consistent with the 
all-particle spectra below tens of PeV. The KASCADE data gives the highest 
energy of the knee, which slightly over-shoots, but is roughly consistent 
with, the all-particle spectra.

\begin{figure*}[!htb]
\centering
\includegraphics[width=1\textwidth]{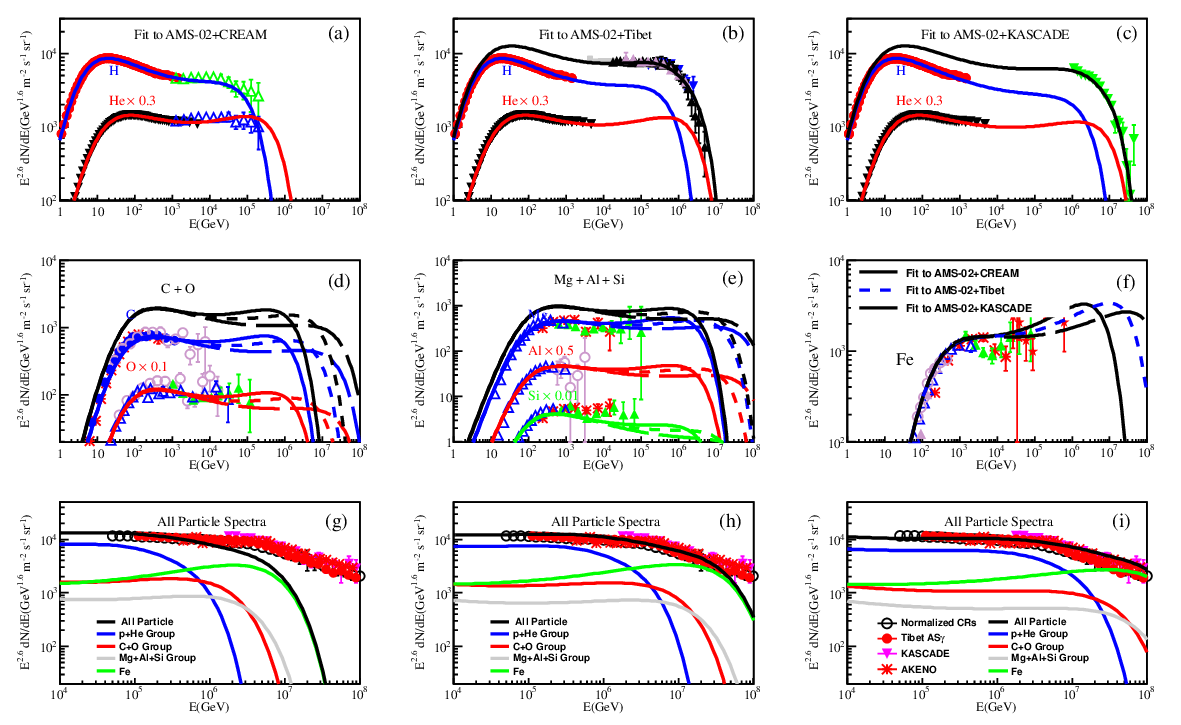}
\caption{Same as Fig. \ref{fig:logEZ} but for the $A$-dependent model.}
\label{fig:logEA}
\end{figure*}

The results for the $A$-dependent scenario are shown in Fig. 
\ref{fig:logEA}. We have similar conclusion with that of the
$Z$-dependent scenario. For the $A$-dependent case, the knee energy 
of protons is smaller by a factor of $\sim2$ compared with that of the
$Z$-dependent case. At present it is difficult to distinguish these
two cases, and we need measurements of the knee of either protons or
Helium to distinguish them.

\section{The spatially-dependent diffusion model}

In the above section, we introduce a ``modified poly-gonato'' model to
reproduce the wide-band spectra of CRs. One possible physical explanation
of the spectral hardening ($\beta \cdot \log E$) is the spatially-dependent
diffusion of particles \citep{2012ApJ...752L..13T,2016ChPhC..40a5101J,
2015PhRvD..92f3001T,2016ApJ...819...54G,2016PhRvD..94l3007F}. 
In a simplified version, i.e., CRs diffuse separately in the disk region 
and halo region (the two-halo model), the hardening of the primary CR 
nuclei and the excesses of secondary particles can be reasonably accounted 
for \citep{2012ApJ...752L..13T,2015PhRvD..92f3001T,2016ApJ...819...54G,
2016PhRvD..94l3007F}. Here we extrapolate this model to the knee region to 
reproduce the results of the phenomenological ``modified poly-gonato'' model.

\subsection{Model description}

We employ the diffusion reacceleration model to describe the propagation
of CR particles \citep[see e.g.,][]{2007ARNPS..57..285S,2017PhRvD..95h3007Y}. 
A cylindrical geometry is assumed. The propagation is confined in a halo 
with half height of $z_h$. The diffusion coefficient, $D_{xx}$, depends 
on both the spatial coordinates $(r,z)$ and the particle rigidity, which 
is parameterized as \citep{2016ApJ...819...54G}
\begin{equation}
D_{xx}(r,z,\rho) = \left \{
    \begin{array}{lll}
    \eta(r,z)\beta\left(\frac{\rho}{\rho_0}\right)^{\varepsilon(r,z)}, &|z|<\xi z_h & ({\rm disk})\\
            D_0\beta\left(\frac{\rho}{\rho_0}\right)^{\delta_0}                 , &|z|>\xi z_h & ({\rm halo})
    \end{array} \right.
\label{Dxx}
\end{equation}
where $\beta$ is the velocity of the particle in unit of light velocity $c$,
$D_0$ represents the normalization of the halo diffusion efficient at
$\rho_0=4$ GV, $\delta_0$ characterizes the rigidity dependence of the
diffusion coefficient, $\xi z_h$ denotes the thickness of the disk,
$\eta(r,z)$ and $\varepsilon(r,z)$ describes the spatial dependence of
the diffusion coefficient in the disk. $\eta(r,z)$ and $\varepsilon(r,z)$
can be related to the source distribution $f(r)$, via a unified form as
\citep{2016ApJ...819...54G},
\scriptsize
\begin{equation}
F(r,z) = 
 \left \{
   \begin{array}{ll}
   \left(1/[1+e^{f(r)}]-A_i\right)[1-(z/\xi z_h)^4]+F_0\cdot(z/\xi z_h)^4 & ({\rm disk})\\
          F_0                        & ({\rm halo})
   \end{array} \right.
\label{Dconstruction}
\end{equation}
\normalsize
where $A_i$ is a constant with $i$ denoting $\eta$ or $\varepsilon$, $F_0$ is
the $D_0$ and $\delta_0$.

The reacceleration is described by a diffusion in the momentum space,
with a relation between $D_{\rm pp}$ and $D_{\rm xx}$ as
\citep{1994ApJ...431..705S}
\begin{equation}
D_{pp}D_{xx} = \frac{4p^2v_A^2}{3\delta(4-\delta ^2)(4-\delta)w},
\end{equation}
where $p$ is the momentum of a particle, $\delta$ is the power-law index
of the rigidity dependence of the spatial diffusion coefficient (see Eq.
(\ref{Dxx})), $v_A$ is the Alfven speed, $w$ is the ratio of the
magnetohydrodynamic wave energy density to the magnetic field energy
density which is assumed to be 1.

The injection spectrum of CR nuclei is assumed to be broken power-law
with an exponential cutoff
\begin{equation}
q_{\rm inj}(E)=q_0\,e^{-E/E_c^i}\times \left \{
   \begin{array}{ll}
   (E/E_{\rm br})^{-\gamma_1}, & E<E_{\rm br}\\
   (E/E_{\rm br})^{-\gamma_2}, & E\ge E_{\rm br}
   \end{array} \right..
   \label{InjecFormula}
\end{equation}
where $q_0$ is the normalization factor, $E_{\rm br}$ is the break energy,
$\gamma_1,\gamma_2$ are the spectral indices below and above $E_{\rm br}$,
and $E_c^i$ characterizes the spectral cutoff around the knee. The relative
abundances of different nuclei are adopted as the default values used in
DRAGON \cite{2008JCAP...10..018E}. Similar as in Sec. 2, we consider
either the $Z$-dependent and $A$-dependent models of the cutoff.

Low energy particles ($E\lesssim10$ GeV/n) will be modulated by solar
activities, showing suppress of their low energy fluxes. We use the
force-field approximation to account for this solar modulation
\citep{1968ApJ...154.1011G}. In this work, The modulation potential $\Phi$
is fixed to be 550 MV for all the nuclei except for B/C whose modulation
potential is adopted as 200 MV.

\subsection{Results}

\subsubsection{Primaries}

We use the numerical code DRAGON to calculate the spatial dependent
diffusion of CRs \citep{2008JCAP...10..018E}. The injection spectral
parameters are given in Tables \ref{InjectSpect}. The parameter $\gamma_2$
differs for each species. They are tuned to fit the data for the major
compositions. And for the less abundant nuclei, we assume the same
difference of $\gamma_2$ from that of protons as in Sec. 2. The full
compilation of $\gamma_2$ are given in the Appendix.

\begin{figure*}[!htb]
\centering
\includegraphics[width=1\textwidth]{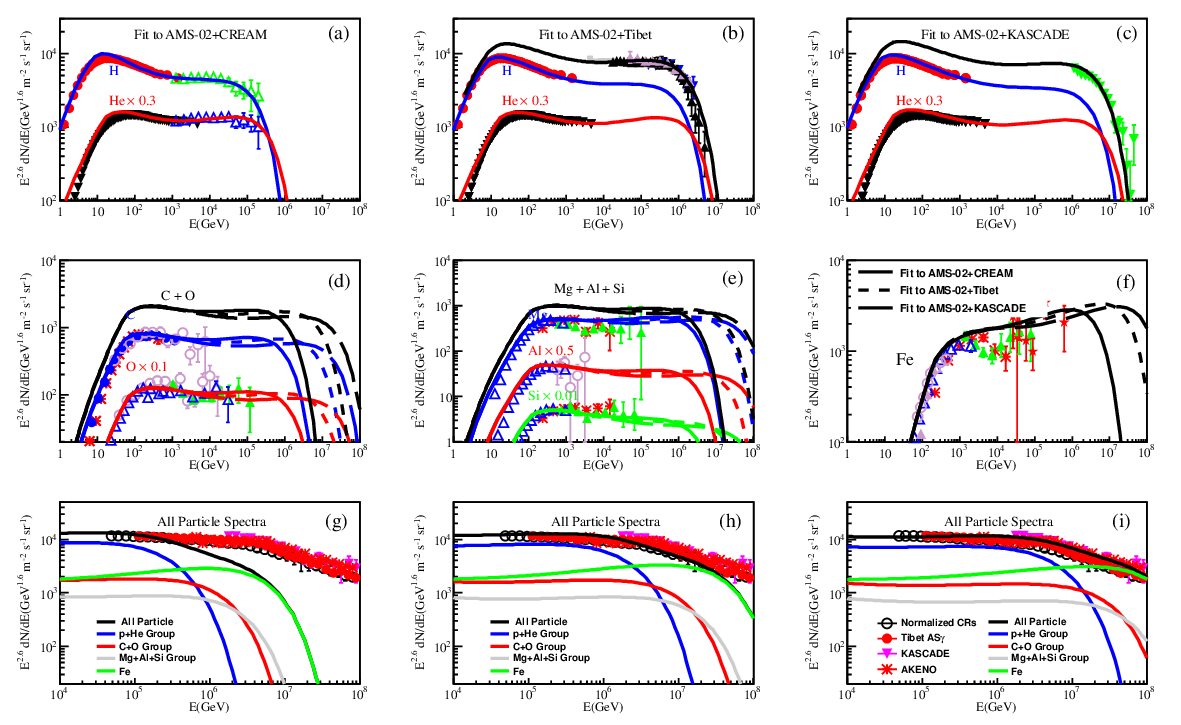}
\caption{Same as Fig. \ref{fig:logEZ}, but for the spatially-dependent
diffusion model.}
\label{fig:propZ}
\end{figure*}

\begin{figure*}[!htb]
\centering
\includegraphics[width=1\textwidth]{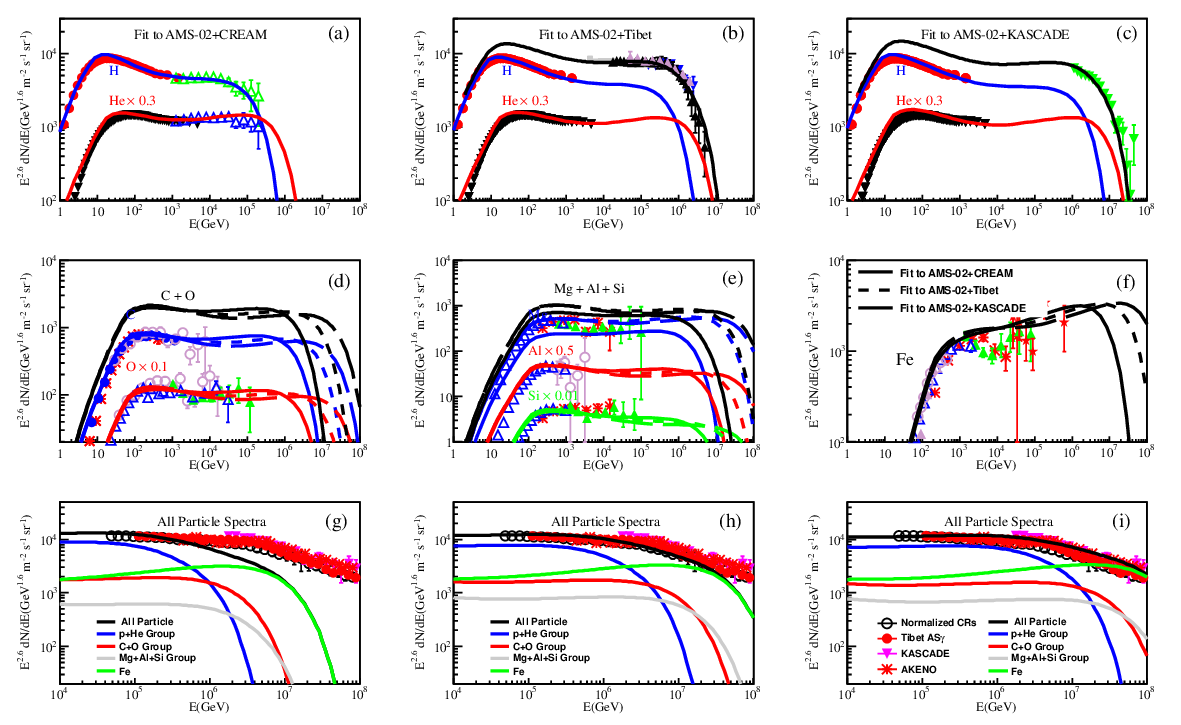}
\caption{Same as Fig. \ref{fig:logEA}, but for the spatially-dependent
diffusion model.}
\label{fig:propA}
\end{figure*}

The propagation parameters are given in Table \ref{paramProp}. 
 Note that, in principle, the allowed parameter space needs to be 
estimated by a global fitting to the data, which is CPU-time consuming
and is left for future studies.
We compare the model predictions to the three data sets of the knee of 
the light components, as described in Sec. 2. Results of the primary CRs 
are shown in Figs.~\ref{fig:propZ} and \ref{fig:propA}, for the $Z$- and 
$A$-dependent cutoff scenarios of the knee. We find that the results are 
very similar to that of the ``modified poly-gonato'' model. It is shown 
that a log-parabolic shape of the energy spectrum is a good approximation 
of a class of models with smooth hardenings.

\begin{table*}[!htb]
\begin{center}
\caption{Injection parameters of the ``spatially-dependent diffusion'' model}
\begin{tabular}{ccccc}\hline \hline
              & mode & AMS-02+CREAM & AMS-02+Tibet & AMS-02+KASCADE    \\
\hline
$E_{br}$(GV)  &       & 9.5       & 9.5            & 9.5   \\
$\gamma_1$    &       & 1.85      & 1.85           & 1.85  \\
$E_c^p$(GeV)  & $Z$   & 1.8$\times10^5$ & 1.1$\times10^6$ & 3.9$\times10^6$ \\
              & $A$   & 1.5$\times10^5$ & 6.6$\times10^5$ & 2.0$\times10^6$ \\
\hline \hline
\end{tabular}
\label{InjectSpect}
\end{center}
\end{table*}

\begin{table*}[!htb]
\begin{center}
\caption{Propagation parameters of the ``spatially-dependent diffusion'' model}
\begin{tabular}{ccccc}\hline \hline
         & AMS-02+CREAM  & AMS-02+Tibet & AMS-02+KASCADE \\
   \hline
   $D_0$ (cm$^2$/s) & $6.8\times 10^{28}$ & $6.8\times 10^{28}$ & $6.8\times 10^{28}$\\
   $\delta_0$    & 0.58                & 0.52                & 0.5 \\
   $v_A$ (km/s)  & 16                  & 16                  & 16   \\
   $z_h$ (kpc)   & 5                   & 5                   & 5    \\
   $\xi $        & 0.14               & 0.12               & 0.11\\
   $A_\eta$      & 0.10                & 0.10                & 0.10  \\
   $A_\varepsilon$& $-0.17$            & $-0.16$             & $-0.14$ \\
\hline \hline
\end{tabular}
\label{paramProp}
\end{center}
\end{table*}

\subsubsection{Secondaries}

\begin{figure*}[!htb]
\centering
\includegraphics[width=0.47\textwidth]{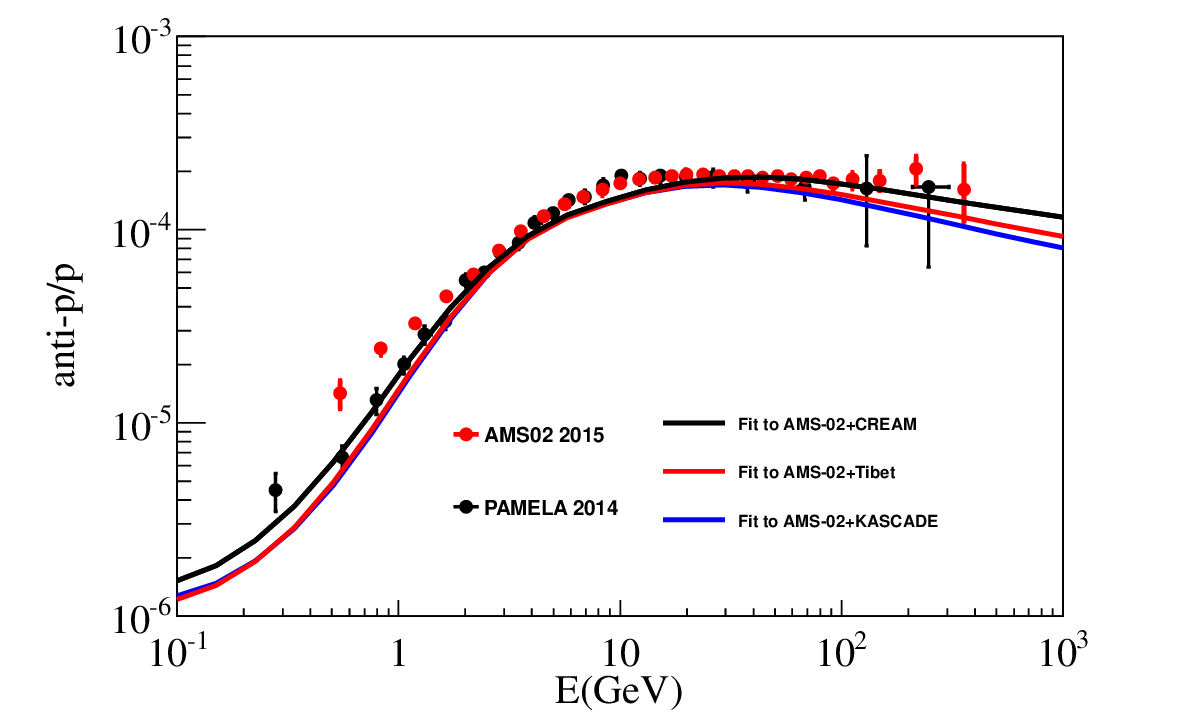}
\includegraphics[width=0.47\textwidth]{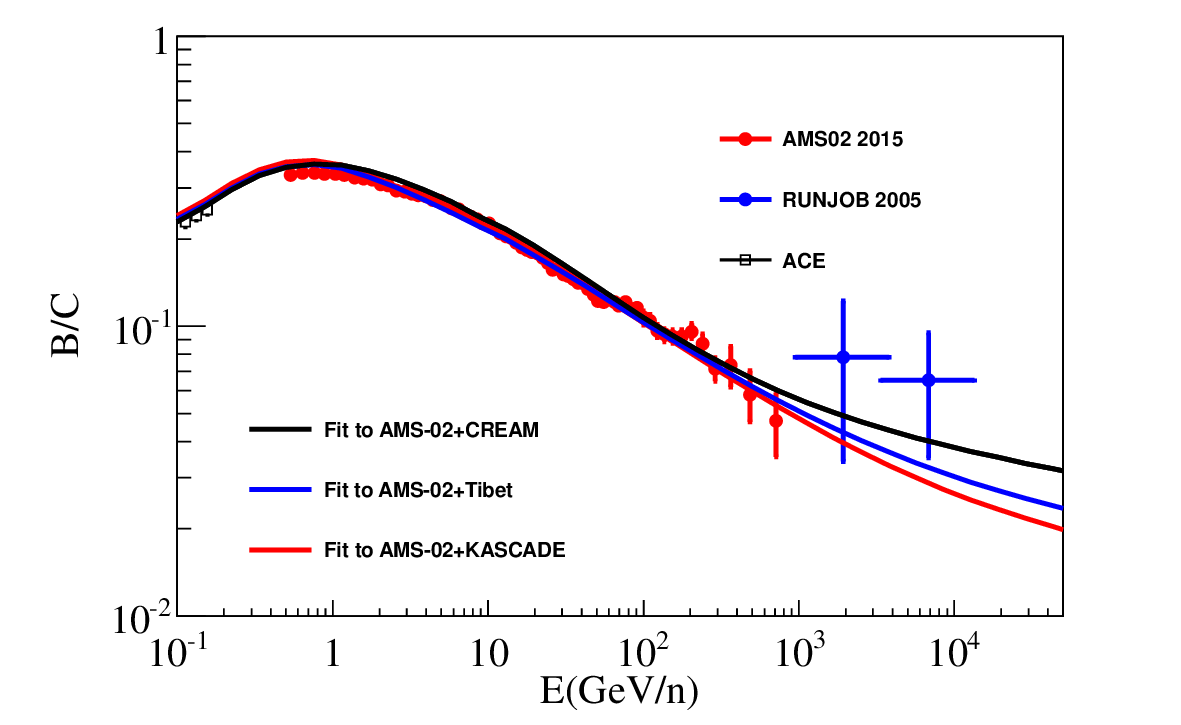}
\caption{The calculated $\bar{p}/p$ (left panel) and B/C (right panel)
ratios of the spatially-dependent diffusion model. The $\bar{p}/p$ data
are from AMS-02 \citep{AMS02-2015} and PAMELA \citep{2014PhysicsReport};
the B/C data are from AMS-02 \citep{AMS02-2015}, ACE
\citep{2000AIPC..528..421D} and RUNJOB \citep{2005ApJ...628L..41D}.}
\label{fig:ratio}
\end{figure*}

Secondary particles will be produced by collisions of primary CRs with
the interstellar medium when they propagate in the Galaxy. It is believed
that most of antiprotons and Borons are such secondaries, which can be
very effective to probe the particle propagation process. We calculate
the expected $\bar{p}/p$ and B/C ratios of this spatially-dependent diffusion
model, as shown in Fig. \ref{fig:ratio}. These results are reasonably
consistent with the observational data. However, we do find that the
second-to-primary ratio becomes asymptotically flatter at high energies,
which is different from the simple uniform diffusion scenario. This can
be tested with future observations of the B/C ratio to higher energies.

\newpage
\subsubsection{Anisotropy}

\begin{figure*}[!htb]
\begin{center}
\includegraphics[width=0.5\textwidth]{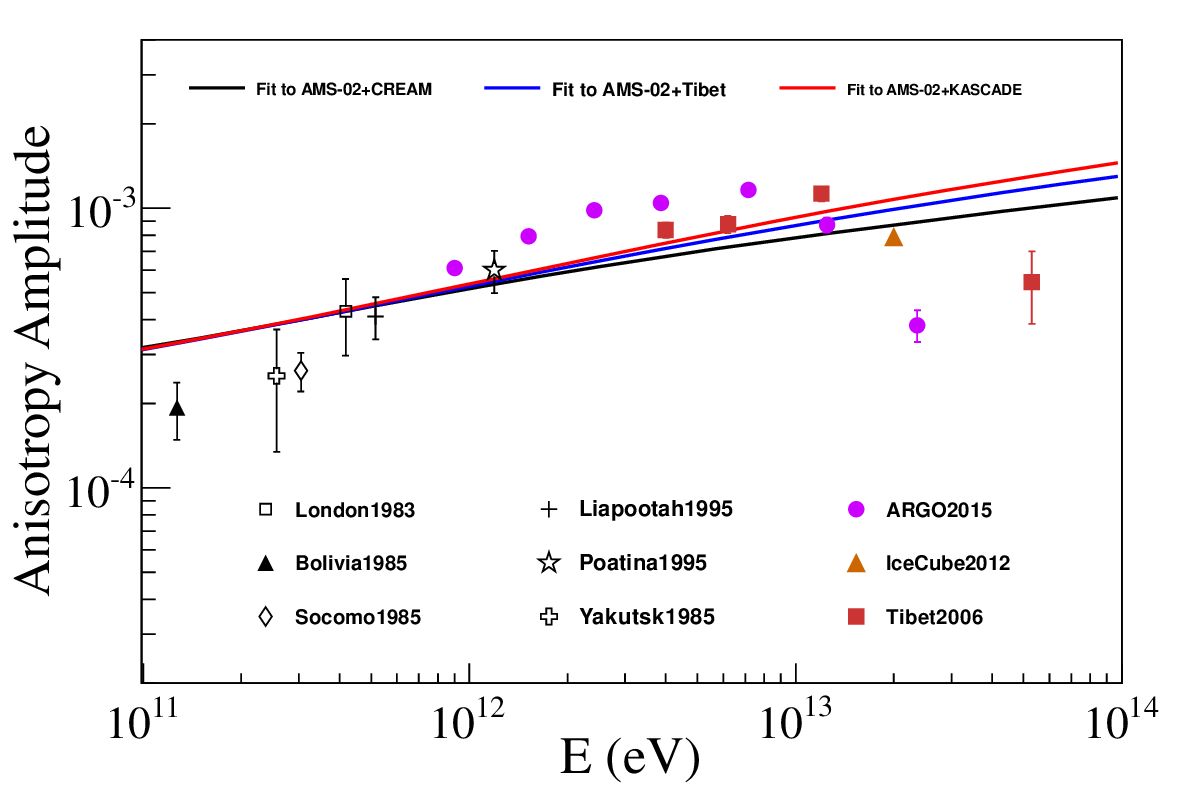}
\caption{The calculated anisotropy of CRs for the spatially-dependent 
diffusion model. The data are from underground muon observations:
London1983 \citep{1983ICRC....3..383T}, 
Bolivia1985 \citep{1985P&SS...33.1069S},
Socorro1985 \citep{1985P&SS...33.1069S}, 
Yakutsk1985 \citep{1985P&SS...33.1069S},
Liapootah1995 \citep{1995ICRC....4..639M}, 
Poatina1995 \citep{1995ICRC....4..635F},
and air shower array experiments:
Tibet2006 \citep{2006Sci...314..439A}, 
IceCube2012 \citep{2012ApJ...746...33A}, 
ARGO2015 \citep{2015ApJ...809...90B}.}
\label{fig:aniso}
\end{center}
\end{figure*}

The flow of CRs will form a dipole anisotropy of arrival directions when
observed at a fixed point. We calculate the amplitude of the dipole
anisotropy of CRs as
\begin{equation}
A=\frac{D}{c}\frac{\nabla\phi}{\phi},
\end{equation}
where $\phi$ is the locally observed differential fluxes of CRs.
The dipole anisotropy amplitude as a function of energy is given in
Fig. \ref{fig:aniso}.
The amplitude of the anisotropy is smaller
than the prediction of the standard diffusion model
\citep{2012ApJ...752L..13T,2015PhRvD..92f3001T,2016ApJ...819...54G,2016PhRvD..94l3007F}, and is consistent
with observations up to a few tens of TeV. Note, however,
the phase of the observed anisotropy shows an evolution with energy,
which can not be simply accounted for by the diffusion process
\citep{ZFeng2015}. More complicated process like the effect of the
local magnetic field and/or local sources may be responsible for it
\citep{2016PhRvL.117o1103A}.

\section{Conclusion and discussion}

Recent observations revealed new features on the CR spectra, including
spectral hardenings at $\sim200$ GV rigidities from balloon or space
detectors and the knee of light components (p and He) from air shower
experiments. In this work we develop a modified version of the
``poly-gonato'' model of the knee, taking into account such new data.
A log-parabolic term of the spectrum is employed to describe the
spectral hardenings. As for the knee, we adopt an exponential cutoff
spectrum to describe it, with the cutoff energy being proportional to
$Z$ or $A$ of each species. We then fit the spectral parameters to
the observational data. Due to the difficulty of the absolute energy
calibration in the air shower experiments, the break positions of the
light component spectra differ to some degree among different experiments.
Therefore the fits are done for different datasets separately,
based on the light component measurements by CREAM, Tibet experiments 
(AS$\gamma$, WFCTA and ARGO-YBJ), and KASCADE. We also try to jointly
fit the CREAM data and the air shower experimental data, and find that
the fitting goodness is relatively poor, expecially for CREAM+KASCADE.
In all the fits, the AMS-02 measurements at low energies ($\lesssim$TeV)
are included.

The results show that the knee energies inferred from different data
groups are marginally consistent with each other. The CREAM data slightly 
favors a relatively low energy knee of protons and Helium nuclei, which 
under-shoots the all-particle spectra individual fits. In this case,
an extra component of CRs below the knee may be required 
\citep[e.g.,][]{2013FrPhy...8..748G,2016A&A...595A..33T}. The Tibet 
experiment and KASCADE data of light components are roughly consistent 
with the all-particle data.

There are no good measurements of CR spectra in the energy range of 
$1-100$ TeV. For example, the proton and Helium spectra by CREAM 
\citep{2017ApJ...839....5Y} differ much from that by ATIC-2
\citep{2007BRASP..71..494P}. A direct comparison of the AMS-02 measured
fluxes of Helium and that by CREAM shows that the CREAM ones are higher by 
about 20\% at TeV/nucleon \citep{2015PhRvL.115u1101A}. Further more precise 
measurements of the energy spectra of various species, by e.g., CALET 
\citep{2007NuPhS.166...43C}, DAMPE \citep{TheDAMPE:2017dtc},
and LHAASO \citep{2010ChPhC..34..249C} will be very important to better
determine the model parameters.

It is also possible that the fitting function, which basically has 
smooth behaviors of the hardening and cutoff, is not good enough to
describe the data. If there are some sharp structures of the spectra,
the CREAM proton and Helium spectra and the all-particle data might be
better consistent with each other. However, in such a case the model may
need fine tuning.

Finally we give a physical model with spatially-dependent diffusion of CRs
to reproduce the results of this phenomenological ``modified poly-gonato''
model. Apart from the potential inconsistencies among different data
sets, the energy spectra of the primary CRs, the secondary-to-primary
ratios, and the amplitude of the anisotropy are shown to be consistent
with observations. This model predicts asymptotical hardening of the
B/C ratio above hundreds of TeV/n, which can also be tested with future
measurements.

\section*{Acknowledgments}
This work is supported by the National Key Research and Development Program
of China (No. 2016YFA0400200), the National Natural Science Foundation
of China (Nos. 11635011, 11761141001, 11663006, 11722328), and the 100 
Talents program of Chinese Academy of Sciences.

\bibliographystyle{unsrt_update}
\bibliography{reco}

\end{multicols}

\section*{Appendix}

We present the spectral parameters of all nuclei up to Iron as used in
this work.

\begin{table}[htp]
\begin{center}
\caption{Low energy spectral parameters of all nuclei of the ``modified
poly-gonato'' model (Sec. 2).}
\begin{tabular}{ccccc}\hline \hline
   $Z$ &$E_{\rm br}/Z$ & $\Phi_0$                      & $\gamma_1$   & $\gamma_2$ \\
       & GV    & (m$^{-2}$s$^{-1}$sr$^{-1}$GeV$^{-1}$) &              &           \\ \hline
   3   & 5.38   & 8.87 $\times 10^{-3}$                 &  0.25       & 2.73      \\
   4   & 5.38   & 5.57 $\times 10^{-3}$                 &  0.25       & 2.94      \\
   5   & 5.38   & 1.56 $\times 10^{-2}$                 &  0.25       & 3.44      \\
   6   & 5.38   & 4.44 $\times 10^{-2}$                 &  0.25       & 2.85      \\
   7   & 5.38   & 7.75 $\times 10^{-3}$                 &  0.25       & 2.91      \\
   8   & 5.38   & 3.12 $\times 10^{-2}$                 &  0.25       & 2.87      \\
   9   & 5.38   & 5.08 $\times 10^{-4}$                 &  0.25       & 2.88      \\
   10  & 5.38   & 3.52 $\times 10^{-3}$                 &  0.25       & 2.83      \\
   11  & 5.38   & 7.09 $\times 10^{-4}$                 &  0.25       & 2.85      \\
   12  & 5.38   & 4.50 $\times 10^{-3}$                 &  0.25       & 2.83      \\
   13  & 5.38   & 8.00 $\times 10^{-4}$                 &  0.25       & 2.85      \\
   14  & 5.38   & 3.50 $\times 10^{-3}$                 &  0.25       & 2.94      \\
   15  & 5.38   & 1.02 $\times 10^{-4}$                 &  0.25       & 2.88      \\
   16  & 5.38   & 4.36 $\times 10^{-4}$                 &  0.25       & 2.74      \\
   17  & 5.38   & 8.81 $\times 10^{-5}$                 &  0.25       & 2.87      \\
   18  & 5.38   & 1.36 $\times 10^{-4}$                 &  0.25       & 2.83      \\
   19  & 5.38   & 1.02 $\times 10^{-4}$                 &  0.25       & 2.84      \\
   20  & 5.38   & 2.30 $\times 10^{-4}$                 &  0.25       & 2.89      \\
   21  & 5.38   & 4.20 $\times 10^{-5}$                 &  0.25       & 2.83      \\
   22  & 5.38   & 1.26 $\times 10^{-4}$                 &  0.25       & 2.80      \\
   23  & 5.38   & 7.00 $\times 10^{-5}$                 &  0.25       & 2.82      \\
   24  & 5.38   & 1.10 $\times 10^{-4}$                 &  0.25       & 2.86      \\
   25  & 5.38   & 1.00 $\times 10^{-4}$                 &  0.25       & 2.71      \\
   26  & 5.38   & 1.05 $\times 10^{-3}$                 &  0.25       & 2.73      \\
\hline \hline
\end{tabular}
\label{polyYuan}
\end{center}
\end{table}

\begin{table}[htp]
\begin{center}
\caption{Injection spectral parameters $\gamma_2$ of all nuclei of the
``spatially-dependent diffusion'' model (Sec. 3).}
\begin{tabular}{ccc}\hline \hline
 Symbol &  $Z$ & $\gamma_2$ \\ \hline
H &   1    & 2.43    \\
He&   2    & 2.36    \\
Li&   3    & 2.26    \\
Be&   4    & 2.47    \\
B &   5    & 2.67    \\
C &   6    & 2.38    \\
N &   7    & 2.44    \\
O &   8    & 2.40    \\
F &   9    & 2.41    \\
Ne&   10   & 2.36    \\
Na&   11   & 2.38    \\
Mg&   12   & 2.36    \\
Al&   13   & 2.38    \\
Si&   14   & 2.47    \\
P &   15   & 2.41    \\
S &   16   & 2.27    \\
Cl&   17   & 2.40    \\
Ar&   18   & 2.36    \\
K &   19   & 2.37    \\
Ca&   20   & 2.42    \\
Sc&   21   & 2.36    \\
Ti&   22   & 2.33    \\
V &   23   & 2.35    \\
Cr&   24   & 2.39    \\
Mu&   25   & 2.18    \\
Fe&   26   & 2.31    \\
\hline \hline
\end{tabular}
\label{propInjt}
\end{center}
\end{table}

\end{document}